\documentclass[sn-mathphys,Numbered]{sn-jnl}

\usepackage{graphicx}%
\usepackage{multirow}%
\usepackage{amsmath,amssymb,amsfonts}%
\usepackage{amsthm}%
\usepackage{siunitx}%
\usepackage{mathrsfs}%
\usepackage[title]{appendix}%
\usepackage{xcolor}%
\usepackage{textcomp}%
\usepackage{manyfoot}%
\usepackage{booktabs}%
\usepackage{algorithm}%
\usepackage{algorithmicx}%
\usepackage{algpseudocode}%
\usepackage{listings}%

\begin{document}

\title{Record-high Anomalous Ettingshausen effect in a micron-sized magnetic Weyl semimetal on-chip cooler  }


\author[1]{\fnm{Mohammadali} \sur{Razeghi}}
\author[1]{\fnm{Jean} \sur{Spièce}}
\author[1]{\fnm{Valentin} \sur{Fonck}}
\author[3,4]{\fnm{Yao} \sur{Zhang}}
\author[1]{\fnm{Michael} \sur{Rohde}}
\author[2]{\fnm{Rikkie} \sur{Joris}}
\author[5]{\fnm{Philip S.} \sur{Dobson}}
\author[5]{\fnm{Jonathan M. R.} \sur{Weaver}}
\author[2]{\fnm{Lino} \sur{da Costa Pereira}}
\author[3,4]{\fnm{Simon} \sur{Granville}}
\author*[1]{\fnm{Pascal} \sur{Gehring}}\email{pascal.gehring@uclouvain.be}

\affil*[1]{\orgdiv{IMCN/NAPS}, \orgname{Université Catholique de Louvain (UCLouvain)}, \orgaddress{\street{Street}, \city{Louvain-la-Neuve}, \postcode{1348}, \country{Belgium}}}

\affil[2]{\orgdiv{QSP}, \orgname{KUL}, \orgaddress{\street{Street}, \city{Leuven}, \postcode{1587}, \country{Belgium}}}

\affil[3]{\orgdiv{Robinson Research Institute}, \orgname{Victoria University of Wellington}, \orgaddress{ \city{Wellington}, \postcode{PO Box 600}, \country{New Zealand}}}

\affil[4]{\orgname{MacDiarmid for Advanced Materials and Nanotechnology}, \orgaddress{ \city{Wellington}, \postcode{PO Box 600}, \country{New Zealand}}}

\affil[5]{\orgdiv{James Watt School of Engineering}, \orgname{University of Glasgow}, \orgaddress{\city{Glasgow}, \postcode{G12 8LT}, \country{United Kingdom}}}

\abstract{Solid-state cooling devices offer compact, quiet, reliable and environmentally friendly solutions that currently rely primarily on the thermoelectric (TE) effect. Despite more than two centuries of research, classical thermoelectric coolers suffer from low efficiency which hampers wider application. In this study, the less researched Anomalous Ettingshausen effect (AEE), a transverse thermoelectric phenomenon, is presented as a new approach for on-chip cooling. This effect can be boosted in materials with non-trivial band topologies as demonstrated in the Heusler alloy $\text{Co}_2\text{MnGa}$. Enabled by the high quality of our material, in situ scanning thermal microscopy experiments reveal a record-breaking anomalous Ettingshausen coefficient of $-2.1$~mV in $\mu$m-sized on-chip cooling devices at room temperature. A significant 44\% of the effect is contributed by the intrinsic topological properties, in particular the Berry curvature of $\text{Co}_2\text{MnGa}$, emphasising the unique potential of magnetic Weyl semimetals for high-performance spot cooling in nanostructures.}

\keywords{Anomalous Ettingshause effect, Magnetic Weyl semimetals, topology, on-chip spot-cooling}

\maketitle

\section*{Main}\label{sec1}

\begin{figure*}[!h]%
\centering
\includegraphics[width=1\textwidth]{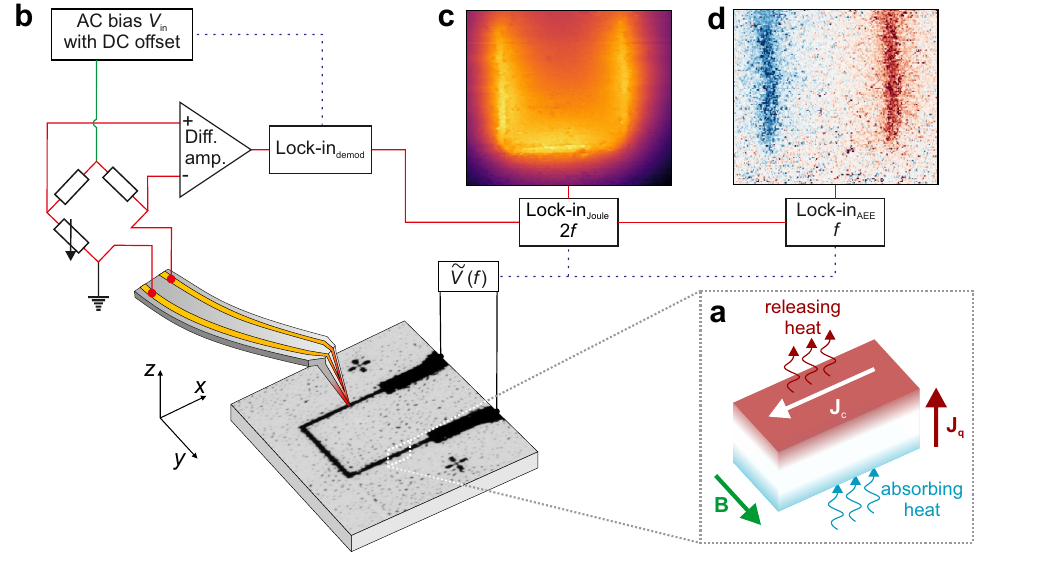}
\caption{\textbf{Schematic of the Scanning Thermal Microscopy experiments used to quantify Ettingshausen cooling}. \textbf{a,} The charge current density $\mathbf{J}_\text{c}$ that flows through an Ettingshausen cooler which is exposed to a magnetic field \textbf{B} induces a transverse heat flux $\mathbf{J}_\text{q}$ which decreases/increases the bottom/top surface temperature. \textbf{b,} An AC current $J_\text{c}$ is injected into the U-shape $\text{Co}_2\text{MnGa}$ device (optical micrograph shown) using an AC voltage bias $\widetilde{V}$ at frequency $f_\text{exc}$ (black lines). The SThM tip is biased with a small AC and DC offset ($V_\text{in}$, green line) using a Wheatstone bridge. A first lock-in demodulates the signal with respect to the tip excitation. Subsequently, a second lock-in demodulates the second ($2\times f_\text{exc}$) and first harmonic ($f_\text{exc}$) corresponding to the Joule (\textbf{c}) and AEE (\textbf{d}) response, respectively. These signals are further converted into temperature variation maps via tip calibration (see methods).}\label{fig1}
\end{figure*}

The constraints that typically limit the efficiency of conventional thermoelectric materials where heat and charge flow in parallel do not apply to the Ettingshausen effect (EE, see Fig. \ref{fig1}b), a \textit{transverse} thermoelectric phenomenon similar to the Hall effect, where heat and charge currents are orthogonal: When a charge current density \textbf{J$_\mathrm{c}$} flows through a material exposed to a magnetic field \textbf{B}, a heat current \textbf{J$_\mathrm{q,EE}$} is generated perpendicularly to both. 
Thanks to their orthogonality, heat and charge flow are decoupled, which eliminates Joule heating's parasitic contributions, significantly boosting conversion efficiency. Consequently, Ettingshausen coolers hold the record for the lowest temperature achieved by room temperature thermoelectric cooling \cite{harman1964oriented}.
However, the Ettingshausen effect typically requires a substantial external magnetic field, often several Tesla, to operate. Recent research has shown that this hurdle can be overcome by leveraging materials with nontrivial band topologies. Akin to the anomalous Hall effect, the thermoelectric counterparts, the anomalous Nernst/Ettingshausen effects, depend on the magnetization of the compound and on the anomalous velocity of electrons linked to the band structure's Berry curvature which can be significant in magnetic materials. Remarkably, a class of quantum materials known as magnetic Weyl semimetals boasts exceptionally large \cite{reichlova2018large} (or even 'giant' \cite{yang2020giant}) anomalous Nernst coefficients due to their topologically non-trivial band structure. This discovery could unlock the high-performance spot cooling potential, as the Nernst and Ettingshausen coefficients are closely linked through the Bridgman relation \cite{bridgman1924connections}. Nonetheless, precise measurement of cooling effects in nanostructures remains challenging due to limitations in instrumentation or material quality \cite{mizuno2022deposition}, and experimental evidence for such significant Ettingshausen cooling efficiency awaits confirmation.
Here, we find a record-high anomalous Ettingshausen coefficient for on-chip coolers with a $\mu$m footprint made of the Heusler alloy $\text{Co}_2\text{MnGa}$. To this end, we performed frequency- and magnetic-field-dependent local thermometry and local thermal transport experiments by means of Scanning Thermal Microscopy (SThM).  Our findings reveal the important contribution of topology on the transverse thermoelectric effect in our samples and demonstrate that microscale anomalous Ettingshausen devices are strong candidates for on-chip cooling applications.

High-quality single-crystalline, epitaxial $\text{Co}_2\text{MnGa}$ thin films with a thickness of 50 nm were deposited on single-crystalline MgO (001) substrates by DC magnetron sputtering \cite{zhang2021berry}. The $\text{Co}_2\text{MnGa}$ films are capped by a thin Ta layer to prevent oxidation (see Methods for details). Since the heat current \textbf{J$_\mathrm{q,AEE}$} generated by the Anomalous Ettingshausen Effect (AEE) depends on the direction of the charge current density $\mathbf{J}_{c}$ and the magnetization \textbf{M} (unit vector \textbf{m}) by \cite{miura2019observation,nagasawa2022anomalous} 

\begin{equation}
\mathbf{J_{\text{q,AEE}}} = \Pi_{\text{AEE}} \cdot \left( \mathbf{J}_{c} \times \mathbf{m} \right),
\label{eq1}
\end{equation} 
where the proportionality constant $\Pi_{\text{AEE}}$ is the Anomalous Ettingshausen coefficient, we patterned the $\text{Co}_2\text{MnGa}$ thin films into microscale U-shaped devices (see Methods). This allows parallel and perpendicular \textbf{m} and $\mathbf{J}_{c}$ to be realised in a single device, leading to areas without ($\mathbf{J}_{c} \times \mathbf{m} = 0$), and with positive ($\mathbf{J}_{c} \times  \mathbf{m}>0$) or negative ($\mathbf{J}_{c} \times \mathbf{m} <0$) Ettingshausen response, respectively. This geometry further allows differentiation between the Ettingshausen effect and other thermoelectric effects like the magneto-Peltier effect \cite{miura2020spin,uchida2018observation} (see Supplementary Fig.~S1).

To measure the local temperature distribution on $\text{Co}_2\text{MnGa}$ on-chip coolers with nanometer spatial and mK temperature resolution, we employed an AC-SThM method \cite{menges2016temperature,harzheim2018geometrically,spiece2018improving} (see Fig.~\ref{fig1}a). To this end, the $\text{Co}_2\text{MnGa}$ U-shaped device is biased by a sinusoidal alternating current (AC) with frequency ${f}$ and amplitude ${J}_{\text{c}}$. Joule heating and Ettingshausen cooling/heating can then be obtained by demodulating the thermometer signal of the SThM probe at $2f$ and $f$, respectively. 

Such Joule and AEE temperature maps recorded at a charge current $I_{\text{c}}=0.952$~mA, $f=17$~Hz, and in-plane magnetic field of $80$~mT are shown in Fig.~\ref{fig1}c,d. While Joule heating raises the temperature of the three sides of the U-shaped device equally, the Ettingshausen effect increases (decreases) the temperature in areas where the magnetic field is perpendicular to the current. For areas where the magnetic field is parallel to the charge current, no temperature change is observed, in accordance with Eq. \ref{eq1}.

\begin{figure*}[!h]%
\centering
\includegraphics[width=1\textwidth]{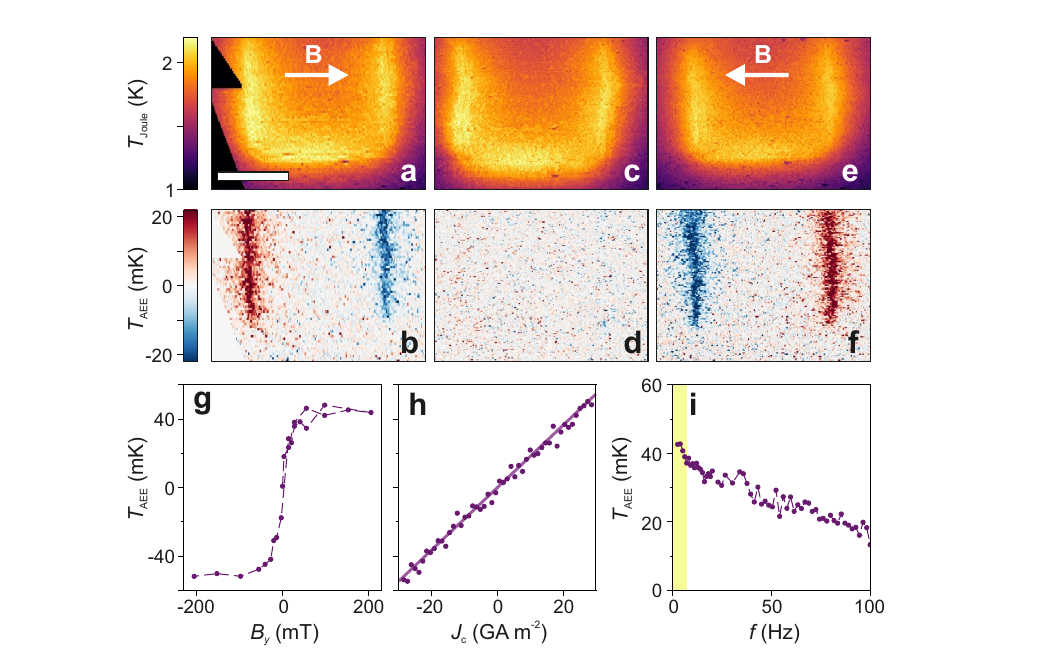}
\caption{\textbf{Anomalous Ettingshausen effect and Joule heating in Co$_2$MnGa.} \textbf{a,c,e,} Maps of $T_{\text{Joule}}$ and \textbf{b,d,f,} $T_{\text{AEE}}$ for the U-shaped 50-nm-thick $\text{Co}_2\text{MnGa}$ film, measured at a frequency of $f$=$17~$Hz, current density $J_\text{c}=19~\mathrm{GA~m^{-2}}$ (current $I_\text{c}=0.95$~mA) and under a magnetic field $B_y$ of $80$~mT (\textbf{a,b}), 0 (\textbf{c,d}), and $-80$~mT (\textbf{e,f}). \textbf{g,} $T_{\text{AEE}}$ as a function of magnetic field $B_{y}$ measured at $J_\text{c}=28.5~\mathrm{GA~m^{-2}}$ and $f=17$~Hz. \textbf{h,} $T_{\text{AEE}}$ as a function of charge current density $J_\text{c}$ measured at $B_y = 80$~mT and $f=17$~Hz. Solid line: linear fit to the data. \textbf{i,} $T_{\text{AEE}}$ as a function of frequency $f$ measured at $J_\text{c}= 19~\mathrm{GA~m^{-2}}$ and $B_y = 80$~mT. Scale bar: 10~$\mu$m.}\label{fig2}
\end{figure*}

In Fig.~\ref{fig2}, we further investigate the magnetic-field dependence of the AEE effect. Figs.~\ref{fig2}a, c, e and b, d, f show the Joule and AEE response, respectively, at -80~mT, 0~mT, and 80~mT. We observe AEE maps of similar spatial distribution and amplitude but with opposite sign for -80~mT and 80~mT as expected from Eq. \ref{eq1}. For $\mathbf{B}=0$ no AEE signal is observed within our experimental resolution limit. The Joule heating maps (Figs.~\ref{fig2}a,c,e) are not affected by the magnetic field. To further investigate the influence of frequency, current density, and magnetic field, we placed the SThM probe at a constant location on the right leg of the U-shaped device. The force between the probe and the sample is kept constant using the AFM feedback loop. Fig.~\ref{fig2}h illustrates the linear correlation between the $T_{\text{AEE}}$ and the applied current density $\mathbf{J}_c$, as expected from Eq. \ref{eq1}, with $\mathrm{d}T_\mathrm{AEE} / \mathrm{d} J_\mathrm{c}=1.8 \times 10^{-12}$ m${}^2$KA${}^{-1}$, while a quadratic response is observed in Joule heating (see Supplementary Fig.~S2). Fig.~\ref{fig2}g shows the magnetic field dependence of the $T_{\text{AEE}}$. We observe a steep increase of the signal at low fields followed by saturation at a field of $B \approx 50$~mT, which qualitatively follows the magnetization of our films measured by SQUID (see Supplementary Fig.~S3c). Such saturation is the hallmark of an anomalous Ettingshausen effect, and the anomalous Ettingshausen coefficient can be extracted from this saturation region. Lastly, we observe that $T_{\text{AEE}}$ decreases with increasing AC frequency of $J_\mathrm{c}$ (Fig.~\ref{fig2}i). This reflects the thermal response time of the system (the U-shaped device and substrate), which is on the order of tens of milliseconds. Conversely, at low frequencies, the signal maximizes and approaches the steady-state value, which is used to quantify $\Pi_{\text{AEE}}$. 

It is worth mentioning that the spatial temperature distribution in the Joule and AEE maps differs noticeably. As shown in the Supplementary Fig.~S3, while Joule heating decays over several $\mu$m around the U pattern, AEE heating/cooling is localized around the device. This directly links to the different origins of the produced heat: While the Joule heating is produced uniformly and isotropically throughout the whole film, the AEE heat current is directional and flows alike from heat source to a heat sink located at the top/bottom sides of the device \cite{das2019systematic}. This confined temperature distribution created by the AEE allows for precise local temperature control that cannot be achieved using conventional methods that have high thermal diffusion. 

\begin{figure}[h]%
\centering
\includegraphics[width=1\textwidth]{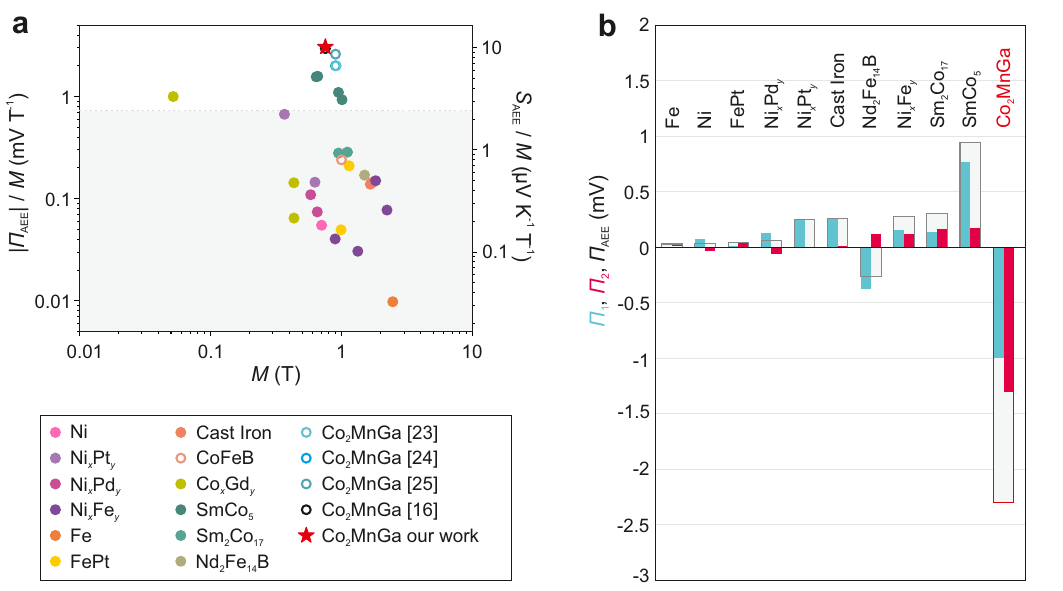}
\caption{Comparison of the Ettingshausen coefficient of Co$_2$MnGa with literature. \textbf{a,} Ettingshausen coefficient normalized by magnetization as a function of magnetization. Filled circles: experimental Ettingshausen data; Empty circles: estimations using experimental Nernst data and the Bridgeman relation: 0.5~mm-thick Fe, Ni, $\text{Ni}_{95}\text{Pt}_{5}$,$\text{Ni}_{75}\text{Pt}_{25}$, $\text{Ni}_{70}\text{Pd}_{30}$, $\text{Ni}_{95}\text{Fe}_{5}$,$\text{Ni}_{75}\text{Fe}_{25}$, $\text{Ni}_{50}\text{Fe}_{50}$, $\text{Ni}_{25}\text{Fe}_{75}$ and  $\text{Fe}_{50}\text{Pt}_{50}$ (ref. \cite{miura2020spin}), 10~nm-thick FePt (ref. \cite{seki2018relationship}), 1~mm-thick ductile and grey cast iron (ref. \cite{nagasawa2022anomalous}), 30~nm-thick $\text{Co}_{86}\text{Gd}_{14}$, $\text{Co}_{78}\text{Gd}_{22}$ and $\text{Co}_{58}\text{Gd}_{42}$ (ref. \cite{seki2019anomalous}), 1.5~mm-thick $\text{Sm}\text{Co}_{5}$, $\text{Sm}_2\text{Co}_{17}$-type magnet and $\text{Nd}_2\text{Fe}_{17}\text{B}$ (ref. \cite{miura2019observation}), 10~nm-thick $\text{CoFeB}$ (ref. \cite{reichlova2018large}), 1.3~mm-thick $\text{Co}_2\text{MnGa}$ (ref. \cite{sakai2018giant}), 0.5~mm-thick $\text{Co}_2\text{MnGa}$ (ref. \cite{guin2019anomalous}), Single crystal $\text{Co}_2\text{MnGa}$ (ref. \cite{xu2020anomalous}), 50~nm-thick $\text{Co}_2\text{MnGa}$ (ref. \cite{hu2022large}). \textbf{b,} Total Ettingshausen coefficient $\Pi_\mathrm{AEE}$ (grey bars) and its two contributions $\Pi_1$ (blue bars) and $\Pi_2$ (red bars). 10~nm-thick Fe, Ni, $\text{Fe}_{50}\text{Pt}_{50}$, $\text{Ni}_{70}\text{pd}_{30}$, $\text{Ni}_{75}\text{Pt}_{25}$ and $\text{Ni}_{50}\text{Fe}_{50}$ \cite{miura2020spin}, 1~mm-thick cast iron \cite{nagasawa2022anomalous}, 1.5~mm-thick $\text{Nd}_2\text{Fe}_{17}\text{B}$, $\text{Sm}\text{Co}_5$ and $\text{Sm}_2\text{Co}_{17}$ \cite{miura2019observation}.}\label{fig4}
\end{figure}

Using Eq. \ref{eq1}, it is possible to derive an expression for the AEE coefficient $\Pi_{\text{AEE}}$ given by experimentally accessible parameters. Considering $T_{\text{AEE}}$ in steady state regime, $\Pi_{\text{AEE}}$ can be expressed as \cite{miura2020high,nagasawa2022anomalous,miura2020spin}:
\begin{equation}
\Pi_{\text{AEE}} = \frac{\kappa \cdot 2 T_{\text{AEE}} }{J_{\text{c}} \cdot L},
 \label{eq2}
\end{equation}
where $\kappa$ is the thermal conductivity and $L$ is the thickness of the film. A factor of 2 is introduced to account for the temperature difference between the top and bottom of the sample. We measured $\kappa$ by SThM (see Supplementary Information Section 6) and find a value of $22\pm 4$ W K\textsuperscript{-1} m\textsuperscript{-1}. Using this value we obtain $\Pi_{\text{AEE}}=-2.1 \pm 0.4$ mV. This agrees well with the value $\approx -2.3$~mV estimated using the Bridgman relation $\Pi_{\text{AEE}}=S_{\text{ANE}}*T$ and the experimental anomalous Nernst coefficient of our films \cite{hu2022large}. In Fig.~\ref{fig4}a we compare $\Pi_{\text{AEE}}$ of $\text{Co}_2\text{MnGa}$ normalized by the magnetization $M$ to values reported in literature (filled shapes). Since only a few studies have experimentally quantified $\Pi_{\text{AEE}}$, we added estimated values for the anomalous Ettingshausen coefficient (hollow shapes) using the experimental Nernst effect coefficient $S_{\text{ANE}}$ and applying the Bridgman relation \cite{seki2018relationship,xu2020finite}. Furthermore, consistent with previous reports on anomalous transport coefficients \cite{miura2020high,miura2020spin,seki2019anomalous}, Fig.~\ref{fig4}a shows that $\Pi_{\text{AEE}}$ of $\text{Co}_2\text{MnGa}$ does not scale with magnetization values (trivial scaling indicated by the gray-shaded area). This suggests the involvement of intrinsic material properties, such as a large Berry curvature resulting from the non-trivial band topology. 

In the following, we will discuss the origin of the large AEE coefficient measured in $\text{Co}_2\text{MnGa}$. The AEE and its Onsager counterpart, the ANE, consist of both intrinsic ($\Pi_{1}$ and $S_{1}$) and extrinsic ($\Pi_{2}$ and $S_{2}$) contributions \cite{miura2019observation, miura2020spin,nagasawa2022anomalous,sumida2020spin}:
\begin{equation}
\Pi_{\text{AEE}} = \Pi_{1} + \Pi_{2}= (S_{1} + S_{2}) T = (\rho_{\text{xx}} * \alpha_{\text{xy}} + \rho_{\text{xy}} * \alpha_{\text{xx}})T.
\end{equation} \label{eq4}
Here, $\rho_{\text{ij}}$ and $\alpha_{\text{ij}}$ are components of resistivity and thermoelectric tensors, respectively. The intrinsic term is a result of the Berry phase which generates an anomalous velocity perpendicular to the applied electric field across occupied states near the Fermi level. The extrinsic term arises from the scattering mechanism induced by effective spin-orbit coupling (SOC) in magnetic materials \cite{nagaosa2010anomalous,miyasato2007crossover}.

To estimate such contributions, we performed magnetoresistance and Hall measurements on our $\text{Co}_2\text{MnGa}$ films (see Supplementary Fig.~S3a and b). Using the experimental values of $S_\text{xx}$, $\Pi_\mathrm{AEE}$, ${\rho_\text{xx}}$ and ${\rho_\text{xy}}$, we calculate a large transverse thermoelectric coefficient $\alpha_{xy}=-2.14~\text{A~(Km)}^{-1}$. This value can be used to estimate $\Pi_{1}$ and $\Pi_{2}$ (see Supporting Information), which we compare to values reported in the literature in Fig. \ref{fig4}b. Notably, $\text{Co}_2\text{MnGa}$ features the record for the highest $\Pi_{\text{AEE}}$ at room temperature. Furthermore, we find that there is a large contribution of 44\% from the Berry curvature $\Pi_1$ to the overall anomalous Ettingshausen effect, underlining the importance of the intrinsic topological nature of $\text{Co}_2\text{MnGa}$ to explain the enormous $\Pi_{\text{AEE}}$ observed in our experiments.

\section*{Conclusion}
In summary, we report an unprecedented anomalous Ettingshausen effect in spot coolers fabricated from Co$_2$MnGa thin films with lateral dimensions of 1 $\mu$m. By combining caloric scanning thermal microscopy measurements and electrical magnetotransport experiments, we uncovered a substantial intrinsic anomalous Ettingshausen component due to the large Berry curvature near the Fermi energy of Co$_2$MnGa. This intrinsic contribution, coupled with the extrinsic counterpart, synergistically enhances the overall cooling efficiency. These results highlight the profound impact of non-trivial band topology on transverse thermoelectric effects and signal transformative potential for a new generation of high-efficiency on-chip coolers. Such advances are poised to address critical needs in thermal management, sensing, healthcare and quantum computing.

\section*{Methods}\label{sec11}

\subsection*{Material synthesis and device preparation}
\subsubsection*{Co$_2$MnGa film growth}

Epitaxial deposition of 50nm $\text{Co}_2\text{MnGa}$ thin films was carried out using a Kurt J Lesker CMS-18 magnetron sputtering system with a base pressure below 3$\times 10^{-8}$ Torr on MgO(001) substrates. Prior to thin film fabrication, substrates were cleaned with Ar plasma and were subsequently annealed at \SI{400}{\degreeCelsius} for 30 minutes in the vacuum chamber. The deposition of $\text{Co}_2\text{MnGa}$ thin films was performed by DC magnetron sputtering from a stoichiometric polycrystalline target at 100~W under 6 mTorr of Ar, with a growth rate of 0.8\AA,s\textsuperscript{-1} at \SI{400}{\degreeCelsius}. During the deposition, the sample holder was rotated, and post-annealing was conducted in situ at \SI{400}{\degreeCelsius} for 20 minutes. Following the cooling down to ambient temperature, a 2~nm Ta protective layer was deposited on the top.

\subsubsection*{Sample Preparation}
Hall bars ($l =2000$~$\mu$m, $w = 150$~$\mu$m) for transport measurements were patterned using photolithography, followed by Ar ion milling. For thermometry measurements, the $\text{Co}_2\text{MnGa}$ is patterned into U-shaped devices using standard electron beam lithography followed by ion milling. The legs have a length of 20~$\mu$m, and a channel width of 1~$\mu$m.

\subsection*{Scanning Thermal Microscopy}
Scanning Thermal Microscopy (SThM) was used for both, temperature mapping and thermal conductivity measurements. Measurements were conducted with a Bruker Dimension Icon Atomic Force Microscope (AFM) under ambient conditions using VITA-DM-GLA-1 probes featuring a palladium heater on a silicon nitride cantilever/tip, with a typical tip radius ranging from 25 to 60~nm. We performed a calibration of the tip radius using a method described elsewhere\cite{spiece2021quantifying,gonzalez2023direct} and obtained 51 nm.The heater, integrated into a modified Wheatstone bridge, was driven by a combined 91~kHz AC and DC bias, as detailed elsewhere \cite{spiece2018improving,Spiece2019}. Detection of the signal was achieved through an SR-830 lock-in amplifier, and the amplitude output was directed to the AFM controller.

\subsubsection*{Temperature measurements}
For magnetic field-dependent AC-thermometry measurements, the sample is biased at a frequency of $f =17$~Hz unless otherwise specified. The SThM tip is driven with $V_\text{pp} =4$~V  and $V_\text{DC} =2$~V at a high frequency of ${f} = 91$~kHz, resulting in a temperature increase of the tip by $\Delta T_\text{tip} \approx 7.5$~K. The direct current (DC) signal is recorded using an SRS830 lock-in at the same frequency as the tip excitation.  $T_\text{AEE}$ and $T_\text{Joule}$ are determined by demodulating the signal using a Zurich Instruments MFLI lock-in amplifier at the 1st and 2nd harmonics while the SThM tip scans over the surface or is held at a constant position. The magnetic field is applied by an home-built electromagnet.

\section*{Data availability}
The data supporting the findings of this study are available in the Article and its Supplementary Information files. Data sets can be found at https://doi.org/xxxx.

\bibliography{sn-bibliography}

\backmatter

\section*{Supplementary information}

Supplementary Figs. 1–6, sections 1-6.

\section*{Acknowledgments}
P.G., J.S. and V.F. acknowledge financial support from the F.R.S.-FNRS of Belgium (FNRS-CQ-1.C044.21-SMARD, FNRS-CDR-J.0068.21-SMARD, FNRS-MIS-F.4523.22-TopoBrain, FNRS-PDR-T.0128.24-ART-MULTI, FNRS-CR-1.B.463.22-MouleFrits, FNRS-FRIA-1.E092.23-TOTEM). P.G. acknowledges financial support from the EU (ERC-StG-10104144-MOUNTAIN), from the Federation Wallonie-Bruxelles through the ARC Grant No. 21/26-116, and from the FWO and FRS-FNRS under the Excellence of Science (EOS) programme (40007563-CONNECT).

\section*{Author information}
\subsection*{Contributions}
P.G. conceived and supervised the study. Y.Z. and S.G. grew the Co$_2$MnGa thin films and performed the magnetotransport measurements. R.J. and L.dC.P. performed the SQUID experiments. V.F. and Mi.R. fabricated the U-shaped devices. P.S.D. and J.M.R.W. fabricated the SThM probes. Mo.R., V.F. and J.S. performed the SThM experiments and evaluated the data. Mo.R., J.S. and P.G. wrote the manuscript with input from all co-authors. All authors have given approval to the final version of the manuscript.

\section*{Declarations}
\subsection*{Conflict of interest}

The authors declare no competing interests.

\end{document}